\documentclass{ieeeconf}
\newcommand{\norm}[1]{\left\lVert#1\right\rVert}
\renewcommand{\H}{\mathcal{H}}
\newcommand{\g}{\tilde{g}}
\newcommand{\gs}{\tilde{g}^*}
\newcommand{\go}{\tilde{g}^\text{opt}}

\newtheorem{definition}{Definition}[section]

\usepackage[utf8]{inputenc} % set input encoding (not needed with XeLaTeX)
\usepackage{amsmath}
\usepackage{amssymb}
\usepackage{algorithm}
\usepackage{algorithmic}
\usepackage{graphicx}
\usepackage{tikz}
\usetikzlibrary{shapes,arrows}

\newlength\figureheight
\newlength\figurewidth
\title{Approximate Regularization Path for Nuclear Norm Based $H_2$ Model Reduction}
\author{Niclas Blomberg, Cristian R. Rojas, and Bo Wahlberg
\thanks{
This work was supported by
the European Research Council under the advanced grant
LEARN, contract 267381and  by the Swedish Research Council under contract 621-2009-4017. The authors are with the Department of Automatic Control and ACCESS Linnaeus Center, School of Electrical Engineering,
KTH--Royal Institute of Technology, SE-100 44 Stockholm, Sweden.
(e-mail: {\tt \footnotesize \{nibl, crro, bo\}@kth.se}.) }
}
\date{} % Activate to display a given date or no date (if empty),
\begin{document}

\maketitle

\begin{abstract} This paper concerns  model reduction of dynamical systems using the nuclear norm of the Hankel matrix to make a trade-off between model fit and model complexity. This results in a convex optimization problem where this trade-off is determined by one crucial design parameter.  The main contribution is a methodology to approximately calculate all solutions up to a certain tolerance to the model reduction problem as a function of the design parameter. This is called the regularization path in sparse estimation and is a very important tool in order to find the appropriate balance between fit and complexity. We extend this to the more complicated nuclear norm case. The key idea is to determine when to exactly calculate the optimal solution using an upper bound based on the so-called duality gap. Hence, by solving a fixed number of optimization problems the whole regularization path up to a given tolerance can be efficiently computed. We illustrate this approach on some numerical examples.
\end{abstract}


\section{Introduction}

The principle of parsimony states that the simplest of two competing theories is to be preferred.  In engineering and science this translates into that a simple model that is good enough for the intended application is preferred compared to a more complex one. Model order reduction concerns methods to find an approximate lower order model of dynamical systems and corresponding error bounds. The advantages of working with lower order models include faster simulation, easier control design and more robust implementations. See \cite{Antoulas:2005} and \cite{Obinata:2001} for references.

Consider a stable scalar discrete dynamical system with transfer function 
\begin{equation}
G_o(z)=\sum_{k=1}^\infty g_{o,k}z^{-k},
\label{eq:1}
\end{equation}
where $\{g_{o,k}\}$ is the impulse response sequence. The model reduction problem is how to find a transfer function 
$$
G(z)=\sum_{k=1}^\infty g_kz^{-k}
$$ of lower order  $n$ such that $G\approx G_o$.

Many approaches to the model order reduction problem have been taken, e.g. balanced truncation \cite{Moore:1981}, Hankel-norm model reduction \cite{Glover:1984}, $L_2$ model reduction \cite{Tjarnstrom-Ljung-02}, and $H_\infty$ model reduction \cite{Zhou-Doyle-Glover-96}.

However, a useful way to measure the approximation error is to use the $H_2$ norm
$$
||G-G_o||_2^2=\sum_{k=1}^\infty [g_k-g_{o,k}]^2.
$$
The corresponding $H_2$ model reduction problem
$$
\min_{G}||G-G_o||_2,\; \mbox{ subject to degree } G=n
$$
is notoriously difficult and many alternative schemes have been proposed. For example, as in Chapter 8 of \cite{Skelton:98}, we can write the constraint as a rank constraint. Our problem will then be a rank minimization problem.

The rank minimization problem is about finding a matrix with minimum rank subject to a set of convex constraints. It has recently been given more and more attention since it appears in many areas, such as control, system identification, and machine learning. However, these problems are in general NP-hard \cite{Mohan:2010}, and many relaxations of them have been explored.

One popular relaxation technique for the rank minimization problem is the nuclear norm minimization heuristic, as discussed in e.g. \cite{Recht:2010}, \cite{Fazel:2001}. It is a convex relaxation which uses the fact that the rank of a matrix follows its nuclear norm (which is defined as the sum of the singular values) in the sense that minimizing the nuclear norm will correspond to minimizing the rank.

Although the nuclear norm minimization problem has been given much attention recently, there are still aspects of it that have not yet been understood. One such aspect is to distinguish cases when the heuristic works and when it does not.

Another aspect concerns choosing the regularization parameter. Although the parameter can often be upper bounded (see \cite{Sadigh:2013}), the particular choice is difficult and has great impact on the trade-off between godness-of-fit and model order. Hence, there is a need to study the impact of the regularization parameter over the whole parameter space.

Following up on this issue we are here interested in outlining the full regularization path in a computationally inexpensive way. Inspired by \cite{Giesen:2012} we suggest an $\varepsilon$-guaranteed regularization path for a specific problem set-up: an $H_2$ minimization problem with a nuclear norm constraint. The idea is to define a certain duality gap that is an upper bound on the approximation error inside which we can confine the true regularization path with a tolerance level $\varepsilon$. 

To map out the full regularization path is useful in many applications. For example, it provides the user with information upon which he/she can select model order. Another use arises when we study iterative re-weighting of the nuclear norm minimization problem, as studied in \cite{Mohan:2010}. Then, the parameter choice can be very tricky since the proper choice can differ from iteration to iteration.

This paper is structured as follows: In Section \ref{problem formulation} we formulate our problem. In Section \ref{method} we introduce a way to approximate the solution to our problem and then establish a bound on the duality gap, which makes it possible to confine the true solution in an $\varepsilon$-approximate region below the approximation. In Section \ref{method} we also suggest an algorithm for implementation of our theory. Finally, we make simulation examples and present our conclusion in Sections \ref{simulation results} and \ref{conclusion}, respectively.

%Model order reduction was first developed in the area of systems and control theory after the success of success projection methods for large scale linear systems. The subject is nowadays studied in diverse applications within systems and control theory as well as in numerical analysis. \cite{Schilders:2008}

%The aim of model order reduction is to find a system approximation of low order compared to an original large scale system. This will reduce computational costs, such as CPU/time and storage, as often desired in applications in industry. However, the low-order approximation needs to preserve the input-output properties of the large scale system with some approximation sufficiency.

\section{Problem Formulation} \label{problem formulation}

Consider a stable scalar discrete dynamical system transfer function as in (\ref{eq:1}) but in a truncated version
\begin{equation} \label{G_o_trunc}
G_o(z)=\sum_{k=1}^{k_\text{max}} g_{o,k}z^{-k},
\end{equation}
where $k_\text{max}$ is assumed to be large enough for the truncated impulse response coefficients to be negligible.
Our aim is to find a low-order approximation $G$ of $G_o$:
\begin{equation} \label{G_trunc}
G(z)=\sum_{k=1}^{k_\text{max}} g_kz^{-k}.
\end{equation}

We define the impulse response vectors corresponding to (\ref{G_o_trunc}) and (\ref{G_trunc}), respectively:
\begin{equation} \label{g_o}
\begin{aligned}
g_o &= \begin{bmatrix}
g_{o,1} & g_{o,2} & \ldots & g_{o,k_\text{max}}
\end{bmatrix}^T \\
g &= \begin{bmatrix}
g_{1} & g_{2} & \ldots & g_{k_\text{max}}
\end{bmatrix}^T
\end{aligned}.
\end{equation}

Consider the following linear operator which creates a matrix with squared Hankel structure:
\begin{equation} \label{hankel}
\mathcal{H}(g) := \begin{bmatrix}
g_1 & g_2 & \cdots & g_n \\
g_2 & g_3 & \cdots & g_{n+1} \\
\vdots & \vdots & \ddots & \vdots \\
g_n & g_{n+1} & \cdots & g_{k_\text{max}}
\end{bmatrix},
\end{equation}

\noindent
where $k_\text{max}=2n-1$ is chosen to be odd. Note that a generalization, which we do not consider here for simplicity, is to define an asymmetric $n\times m$ Hankel structure.

We know from linear system realization theory that for a system with impulse response vector $g$ the system order is equivalent to the rank of $\mathcal{H}(g)$, \cite{Kailath-80}. This sheds some light on why Hankel matrix rank minimization plays a central role in model order reduction.

A common and often successful surrogate heuristic for rank minimization is nuclear norm minimization. This is a convex relaxation of the rank minimization problem. The nuclear norm of a matrix $X$ is defined as
\begin{equation*}
\norm{X}_* = \sum\limits_j \sigma_j,
\end{equation*}

\noindent
where $\sigma_j$ are the singular values of $X$.

\subsection{Problem Statement}

The regularized nuclear norm minimization problem has been presented in various forms in the literature; the reader can compare \cite{Fazel-Hindi-Boyd-01}, \cite{Recht:2010}, and \cite{Gleich:2011}. The nuclear norm penalty is often seen in the objective function but here we state an equivalent version with a nuclear norm constraint.

Here we are interested in an $H_2$ cost, since it is a useful way to measure the approximation error. For $g_o$, $g$, and $\mathcal{H}(g)$ as in ($\ref{g_o}$) and ($\ref{hankel}$) we formulate the following regularized $H_2$ model reduction problem:
\begin{equation} \label{pre-problem}
\begin{aligned}
& \underset{g}{\text{minimize}}
& & \|g-g_o\|_2^2 \\
& \text{subject to}
& & \|\mathcal{H}(g)\|_* \leq t,
\end{aligned}
\end{equation}

\noindent
where $t$ is the regularization parameter. A sufficiently large value of $t$ will give a perfect fit, $g=g_o$, while small $t$ give lower rank of the system.

We comment here as a motivation for future work that the cost in (\ref{pre-problem}) may be extended to a weighted version. With appropriate weights we could then use the maximum likelihood approach for model reduction defined in \cite{Wahlberg:1989}.

In order to get rid of the regularization parameter in the constraint we reformulate the problem in ($\ref{pre-problem}$) to an equivalent version. With $g_o$, $g$, and $\mathcal{H}(g)$ as in ($\ref{g_o}$) and ($\ref{hankel}$) the reformulated version of Problem (\ref{pre-problem}) is
\begin{equation} \label{problem}
\begin{aligned}
& \underset{\tilde{g}}{\text{minimize}}
& & \|t\tilde{g}-g_o\|_2^2 \\
& \text{subject to}
& & \|\mathcal{H}(\tilde{g})\|_* \leq 1,
\end{aligned}
\end{equation}

\noindent
where, again, $t$ is the regularization parameter and $\tilde{g}=\frac{g}{t}$.

We also introduce the following notion of the objective function:
\begin{equation} \label{obj}
f_t(\tilde{g}) = \|t\tilde{g}-g_o\|_2^2.
\end{equation}

\section{Method} \label{method}

Our approach follows the one in \cite{Giesen:2012}, which we specialize to our problem set-up.
\vskip\baselineskip
\fbox{\parbox{220 pt}{Here is an outline of the idea: We want to solve Problem (\ref{problem}) only for a sparse set of points along the regularization path. We call these points $t_i^*, i=1,\ldots,m$, for some $m$. When we have solved Problem (\ref{problem}) in one such point $t^*$ we decide to approximate the solution in some region $t>t^*$. Eventually, the approximation will diverge too far from the true solution, so we decide to stop and re-solve Problem (\ref{problem}).}}
\vskip\baselineskip
The following approximation of $f_t(\tilde{g})$ (defined in (\ref{obj})) is used in the region $t>t^*$:
\begin{equation} \label{appr}
f_t(\tilde{g}^\text{opt}_t) \approx f_t(\tilde{g}^*),
\end{equation}

\noindent
where $\tilde{g}^*$ and $\tilde{g}^\text{opt}_t$ are optimal solutions to Problem (\ref{problem}) in $t=t^*$ and $t>t^*$, respectively. This means that $\tilde{g}^*$ is kept fixed and ($\ref{obj}$) is evaluated for $t>t^*$.

\subsection{The Duality Gap}

The next issue is to decide at which point (when increasing $t$) to stop approximating and instead re-solve Problem (\ref{problem}). To do this will define an upper bound on the approximation error. When this upper bound reaches a certain tolerance level $\varepsilon$, we stop and recompute. In resemblance with \cite{Giesen:2012} we can call the upper bound the {\it duality gap}. It is an upper bound on the approximation error.

To compute an upper bound on the approximation error
$$
f_t(\tilde{g}^*) - f_t(\tilde{g}^\text{opt}_t),
$$
where $\tilde{g}^*$ and $\tilde{g}^\text{opt}_t$ are optimal solutions to Problem (\ref{problem}) in $t^*<t$ and $t$, respectively, we need a lower bound on $f_t(\tilde{g}^\text{opt}_t)$. To this end, let
$$
C = \left\{ \begin{array}{r l}
\underset{\tilde{g}}{\text{min}} & \|\tilde{g}-\frac{1}{t}g_o\|_2^2 \\
\text{s.t.} & \|\mathcal{H}(\tilde{g})\|_* \leq 1,
\end{array} \right\},
$$
which corresponds to the optimal cost of Problem (\ref{problem}) divided by $t^2$.

Now, we try to relax the constraint. From the subdifferential of the nuclear norm (see \cite{Recht:2010}) we get that
\begin{equation} \label{W}
\norm{\H(\tilde{g})}_* \geq \norm{\H(\tilde{g}^*)}_* + \langle UV^T+W, \sum\limits_k H_k(\tilde{g}_k - \tilde{g}_k^*) \rangle
\end{equation}
where $\tilde{g}^*$ solves Problem (\ref{problem}) for a particular $t=t^*$, $\langle \cdot \rangle$ is the standard inner product, $U\Sigma V^T = \H(\tilde{g}^*)$ is a compact singular value decomposition, $W$ is any $n\times n$-matrix obeying $\norm{W}\leq 1$ and $U^TW=WV=0$, and $H_k$ is the Hankel matrix (see (\ref{hankel})) of a vector with zeros everywhere except at the $k^\text{th}$ element which is one.

We rewrite
$$\langle UV^T+W, \sum\limits_k H_k(\tilde{g}_k - \tilde{g}_k^*) \rangle = h^T(\g-\gs),$$
where we have defined the vector $h$ with elements $h_k = \text{tr}[H_k(UV^T+W)]$. Then, the constraint $\norm{\H(\g)}_*\leq 1$ can be relaxed to 
$$
\norm{\H(\gs)}_* + h^T(\g-\gs) \leq 1,
$$
or
$$
h^T(\g-\gs) \leq 0
$$
since $\norm{\H(\gs)}_*=1$ due to the optimality of $\tilde{g}^*$. This relaxation gives
$$
C \geq \left\{ \begin{array}{r l}
\underset{\tilde{g}}{\text{min}} & \|\tilde{g}-\frac{1}{t}g_o\|_2^2 \\
\text{s.t.} & h^T(\g-\gs) \leq 0,
\end{array} \right\},
$$
where the optimal solution to the right hand side can be explicitly computed by the projection theorem: take $\g^\text{opt}=\frac{1}{t}g_o+\alpha h$, where $\alpha$ has to be chosen such that
$$
h^T\left(\frac{1}{t}g_o+\alpha h-\gs \right)=0.
$$
This gives
$$
\alpha = - \frac{1}{\norm{h}_2^2}\left( \frac{1}{t}h^Tg_o - h^T\gs \right),
$$
or
\begin{align*}
\go = \frac{1}{t}g_o-\frac{1}{\norm{h}_2^2}\left( \frac{1}{t}h^Tg_o - h^T\gs \right)h \\
%&= \frac{1}{t}\left( I - \frac{1}{\norm{h}_2^2}hh^T \right)g_o + \frac{1}{\norm{h}_2^2}hh^T\gs,
\end{align*}
so that the lower bound on $C$ becomes
\begin{align*}
C \geq \|\go-\frac{1}{t}g_o\|_2^2 = \frac{1}{\norm{h}_2^4}\norm{hh^T\left( \gs - \frac{1}{t}g_o \right)}_2^2.
\end{align*}

We have now established that
\begin{align*}
f_t(\tilde{g}^*)& - f_t(\tilde{g}^\text{opt}_t) \\
& \leq \norm{t\gs-g_o}_2^2 - \frac{t^2}{\norm{h}_2^4}\norm{hh^T\left( \gs - \frac{1}{t}g_o \right)}_2^2
\end{align*}
and we can define the following duality gap:

\begin{definition} \label{def:dg}
Let $\tilde{g}^*$ be the argument that solves Problem (\ref{problem}) in $t^*$. Then, for any $t\geq t^*$ the duality gap is defined as
\begin{equation} \label{dg}
\begin{aligned}
d&_t^\text{max}(\g) := \\
& \norm{t\gs-g_o}_2^2 - \frac{1}{\norm{h}_2^4}\norm{hh^T\left( t\gs - g_o \right)}_2^2.
\end{aligned}
\end{equation}
\end{definition}
\vskip\baselineskip

Notice that the duality gap equals zero for $t=t^*$, since, for that value of $t$, $\gs$ is the optimal solution to Problem (\ref{problem}), which implies that the 'error vector' $t^*\gs - g_o$ is orthogonal to the supporting hyperplane $\{\g|h^T\g = 0\}$. In other words, $(t^*\gs - g_o)||h$, which implies that
\begin{align*}
d_{t^*}^\text{max}(\gs) &= \norm{t^*\gs-g_o}_2^2 - \frac{1}{\norm{h}_2^4}\norm{hh^T\left( t\gs - g_o \right)}_2^2 \\
&= \norm{t^*\gs-g_o}_2^2 - \norm{\frac{hh^T}{\norm{h}_2^2}\left( t\gs - g_o \right)}_2^2 \\
&= 0.
\end{align*}

With the definition of the duality gap we have established an upper bound on the approximation error. Indeed, we can confine the optimal solution to Problem (\ref{problem}) at any $t \geq t^*$ to lie within the interval
$$
[f_t(\tilde{g}^*) - d_t^\text{max}(\tilde{g}^*), f_t(\tilde{g}^*)].
$$

To confine the duality gap within a certain tolerance, we introduce the notion of an $\varepsilon$-approximation as in \cite{Giesen:2012}.

\begin{definition} \label{eps-appr}
Let $\varepsilon>0$. Consider an argument $\tilde{g}^*$ that solves Problem (\ref{problem}) for parameter value $t^*$. Then, for any parameter value $t>t^*$, we call any $g = t\tilde{g}$ that is feasible for Problem (\ref{pre-problem}) an $\varepsilon$-approximation if it holds for the duality gap that
\begin{equation} \label{d_leq_eps}
d_t^\text{max}(\tilde{g}^*) \leq \varepsilon.
\end{equation}
\end{definition}

\subsection{Upper Bound on $t$}

We here confine the parameter space for $t$ to an interval $[0, t_\text{max}]$. For a sufficiently large $t$ the feasible set of Problem (\ref{problem}) will contain $\tilde{g}=\frac{g_o}{t}$ and we get $f_t(\frac{g_o}{t})=0$. $t_\text{max}$ is the smallest $t$ that satisfies this, i.e.
\begin{equation} \label{t_max}
t_\text{max} = \norm{\mathcal{H}(g_o)}_*.
\end{equation}

We note that other bounds occur in other versions of the nuclear norm minimization problem. There is unfortunately no simple connection between these. In \cite{Sadigh:2013}, where the nuclear norm is a penalize term in the objective function, the calculation of the bound involves the subdifferential of the nuclear norm, giving a more involved derivation of the parameter bound.

\subsection{Algorithm}

The above results give not only explicit bounds on the approximation error, but also suggests a straightforward implementation. Our algorithm is outlined in Algorithm 1.

We sketch the procedure as follows: Let $i=0,1,\ldots,m$, where $m$ is a yet unknown integer representing the number of times we solve Problem (\ref{problem}) along the regularization path. Consider a solution $f_{t_i^*}(\tilde{g}_i^*)$ that solves Problem (\ref{problem}) in $t_i^*$. For each $i$, we record $f_{t_i^*}(\tilde{g}_i^*)$ and $\Sigma_i$, which is the diagonal singular value matrix of $\mathcal{H}(g_i^*)$, where $g_i^* = t_i^* \tilde{g}_i^*$. We then calculate the subsequent point $t_{i+1}^*$ by solving (compare (\ref{d_leq_eps}))
$$d_{t^*_{i+1}}^\text{max}(\tilde{g}^*_i) = \varepsilon,$$

\noindent
where the duality gap $d_t^\text{max}(\tilde{g}^*)$ is defined in (\ref{dg}). For this subsequent point $t_{i+1}$ we will again solve Problem (\ref{problem}). Next, we approximate the solution path in the region $t_i^*<t<t_{i+1}^*$ using (\ref{appr}). The procedure iterates from $t=0$ up to $t_\text{max}$, where $t_\text{max}$ is defined in ($\ref{t_max}$).

\begin{algorithm}                    % enter the algorithm environment
\begin{algorithmic}                    % enter the algorithmic environment
	\STATE \textbf{Algorithm 1. }
	\STATE \textbf{Input: } $\varepsilon, t_\text{max}$, function $f_t(\tilde{g})$ defined by ($\ref{obj}$)\\
	\STATE \textbf{Output: } Consider $i=1,\ldots,m$.
	\STATE $\{f_{t_i^*}(\tilde{g}_i^*)\}$: a set of exact solutions to Problem (\ref{problem}) \\
	\STATE $\{f_t(\tilde{g}_i^*)\}$: approximate solutions in $(t_{i-1}^*,t_{i}^*)$
	\STATE $\{\Sigma_i\}$: the singular values of $\mathcal{H}(g_i^*)$
	\STATE
    \STATE $i = 0$
    \STATE $t_i^* = 0$
    \STATE $\tilde{g}_i^* = 0$
    \WHILE{$t_i^* \leq t_\text{max}$}
        \STATE $i = i + 1$ \\
        \STATE $t_i^* = $ solution to $d_{t_i^*}^\text{max}(\tilde{g}^*_{i-1})=\varepsilon$\\
        \STATE $\tilde{g}_i^* = \underset{\|\mathcal{H}(\tilde{g})\|_* \leq 1}{\text{arg min}} f_{t_i^*}(\tilde{g})$
        \STATE $\Sigma_i = $ singular values of $\mathcal{H}(g_i^*)$
        \FOR{$k = \text{linspace}(t_{i-1}^*,t_i^*)$}
        	\STATE evaluate $f_k(\tilde{g}_{i-1})$
        \ENDFOR
    \ENDWHILE
\end{algorithmic}
\end{algorithm}

\section{Simulation Results} \label{simulation results}

In this section we show the results of implementing Algorithm 1 for two different set-ups $\{g_o,k_\text{max},\varepsilon\}$. In the first case (see Figure 1) we have chosen a $6^\text{th}$ order system with four relatively small singular values, i.e.~it can be approximated by a $2^\text{nd}$ order system. In the second case (see Figure 2), we have chosen a $100^\text{th}$ order system with around ninety relatively small singular values, i.e.~it can be approximated by a $10^\text{th}$ order system. In both cases, $k_\text{max}$ is chosen large enough for the truncated impulse responses to be negligible. Throughout the simulations we have chosen $W=0$ in (\ref{W}) for convenience.

Figures 1 (a) and 2 (a) show a shaded/green area enclosed by $f_t(\tilde{g}^*)$ (the approximate path) from above and $f_t(\tilde{g}^*) - d_t^\text{max}$ from below. (The notation here might be confusing since $\tilde{g}^*$ is different for each subinterval $(t_{i-1}, t_i), i=1,\ldots,m$ along the regularization path.) For the black, dashed lines we have solved Problem (\ref{problem}) for a dense grid of $t$ and it can hence be said to represent the true regularization path. The black vertical lines indicate the stopping points where we have re-solved Problem (\ref{problem}).

The calculation of the duality gap suffers from some numerical errors, which are most significant for very small values of $t$; hence the $t$-axis does not start from zero. These numerical errors result in that the gap is not always zero in $t^*_i$ (the point where we solve Problem (\ref{problem}) exactly). They can to some extent be explained by the division of $t$ in (\ref{dg}). Further, they are certainly explained by that we truncate the matrices $U$ and $V$ defined in (\ref{W}). The truncation is necessary since otherwise $U$ and $V$ will always have full rank due to numerical rounding and it can be verified that (\ref{W}) does not make sense.

Figures 1 (b) and 2 (b) plot singular values of $\mathcal{H}(g_i^*)$, where $g_i^* = t_i^* \tilde{g}_i^*$ and $i=1,\ldots,m$. We have included only singular values that are of interest in our examples. For example, for the system in Figure 1 (b), where we have used a $6^\text{th}$ order system which can be approximated by a $2^\text{nd}$ order system, we include the $3^\text{rd}$ to $6^\text{th}$ singular values. In Figure 2 (b) we have excluded singular values below $\sigma_6$, which are negligible.

In Figure 1 (b) we see that for $t<0.6$ we get a system of $2^\text{nd}$ order, since we have a drop there in $\sigma_3$; the $3^\text{rd}$ singular value. In Figure 2 (b) we see several drops, and the user has to decide on what model order is desired and when that is achieved.

\begin{figure}[h]
\includegraphics[scale=0.6]{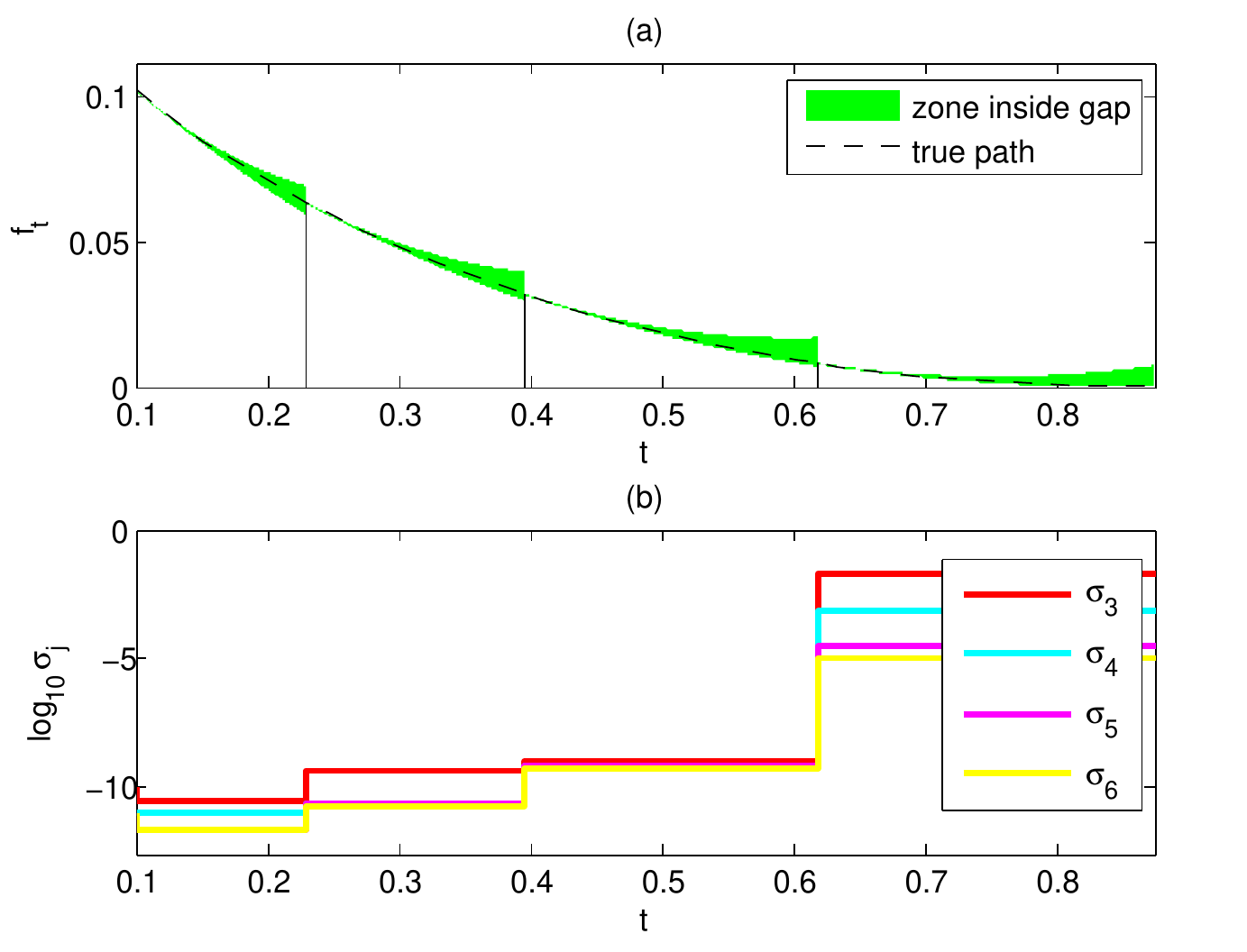}
\label{sys4_2}
\caption{(a) Black dashed: true path, green/shaded area: approximation (upper edge), approximation minus duality gap (lower edge). $(\varepsilon, k_\text{max}, t_\text{max})=(0.01,31,0.87)$ (b) Some relevant singular values of $\H(g^*_i)$ evaluated in $t_i^*,i=1,\ldots,5$.}
\end{figure}

\begin{figure}[h]
\includegraphics[scale=0.6]{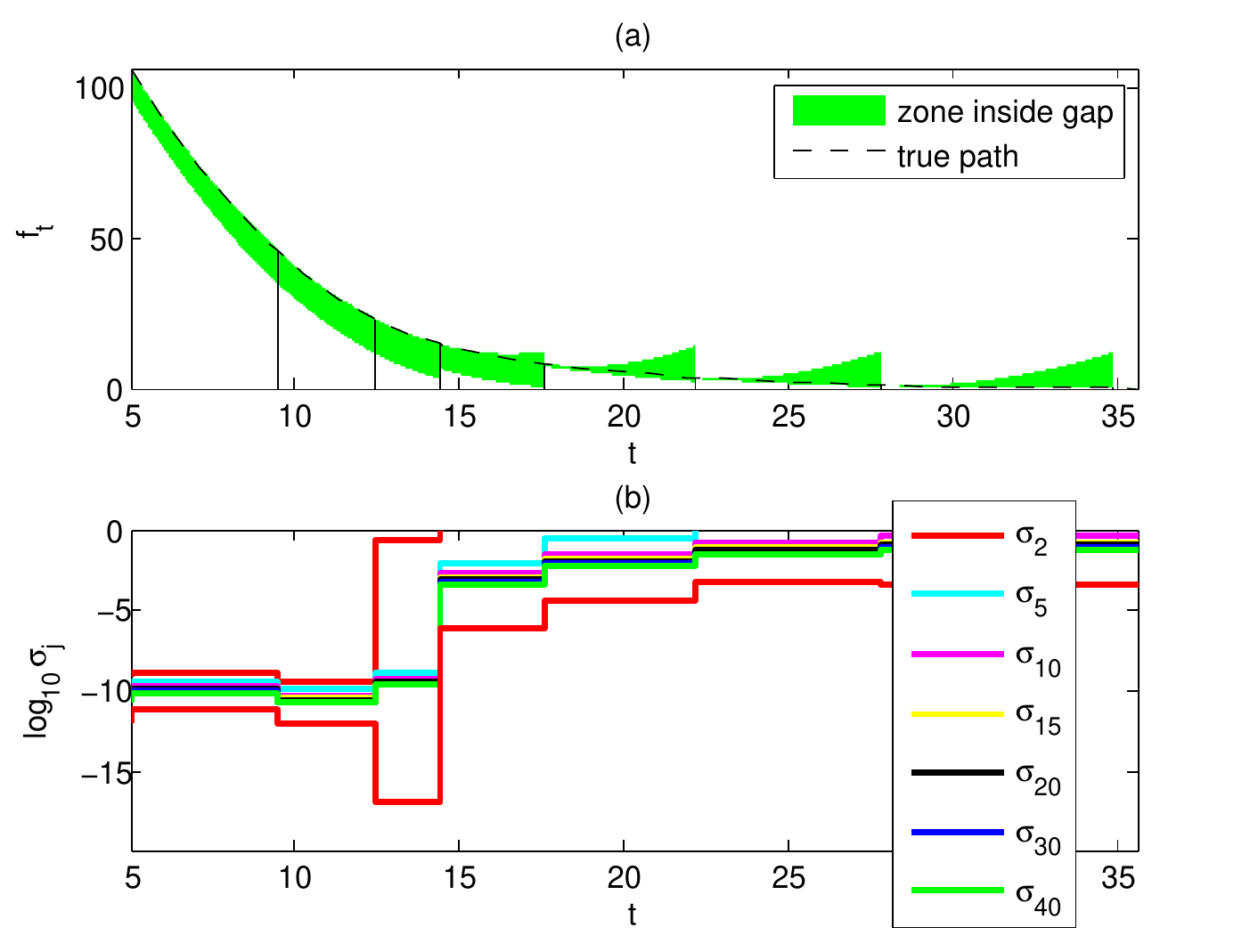}
\label{sys6_2}
\caption{(a) Black dashed: true path, green/shaded area: approximation (upper edge), approximation minus duality gap (lower edge). $(\varepsilon, k_\text{max}, t_\text{max})=(12,51,35.6)$ (b) Some relevant singular values of $\H(g^*_i)$ evaluated in $t_i^*,i=1,\ldots,8$.}
\end{figure}

\section{Conclusion} \label{conclusion}

With this paper we have suggested a method to study Problem (\ref{pre-problem}) over the whole regularization parameter space, inspired by the work in \cite{Giesen:2012}. The simulation result is promising in showing a computationally cheap, approximate regularization path.

This approximate path outlines the effect of the parameter value. The user can then make efficient model order selection. The use of an approximate path also arises e.g. when performing iteratively re-weighted nuclear norm minimization. Then, the outlined path makes it possible to re-choose parameter value in each iteration.

As for future scopes, we aim to explore other versions of our cost function, possibly weighted versions of it. Another extension can be to include input-output data in the problem set-up, turning the problem into a subspace identification problem.

\bibliography{niclas}

\end{document}